
%
%
\input harvmac %
%
%
%
%
%
%
%
%
%
\newif\ifdraft

\noblackbox
\catcode`\@=11
\newif\iffrontpage
%
\ifx\answ\bigans
\def\titleft{\titsm}
\magnification=1200\baselineskip=15pt plus 2pt minus 1pt
%
\advance\hoffset by-0.075truein
\hsize=6.15truein\vsize=600.truept\hsbody=\hsize\hstitle=\hsize
\else\let\lr=L
\def\titleft{\titla}
\magnification=1000\baselineskip=14pt plus 2pt minus 1pt
%
\vsize=6.5truein
\hstitle=8truein\hsbody=4.75truein
\fullhsize=10truein\hsize=\hsbody
\fi
\parskip=4pt plus 10pt minus 4pt

\font\titla=cmr10 scaled\magstep3
\font\tenmss=cmss10
\font\absmss=cmss10 scaled\magstep1
\newfam\mssfam
\font\footrm=cmr8  \font\footrms=cmr5
\font\footrmss=cmr5   \font\footi=cmmi8
\font\footis=cmmi5   \font\footiss=cmmi5
\font\footsy=cmsy8   \font\footsys=cmsy5
\font\footsyss=cmsy5   \font\footbf=cmbx8
\font\footmss=cmss8
\def\footfont{\def\rm{\fam0\footrm}
\textfont0=\footrm \scriptfont0=\footrms
\scriptscriptfont0=\footrmss
\textfont1=\footi \scriptfont1=\footis
\scriptscriptfont1=\footiss
\textfont2=\footsy \scriptfont2=\footsys
\scriptscriptfont2=\footsyss
\textfont\itfam=\footi \def\it{\fam\itfam\footi}
\textfont\mssfam=\footmss \def\mss{\fam\mssfam\footmss}
\textfont\bffam=\footbf \def\bf{\fam\bffam\footbf} \rm}
\def\tenpoint{\def\rm{\fam0\tenrm}
\textfont0=\tenrm \scriptfont0=\sevenrm
\scriptscriptfont0=\fiverm
\textfont1=\teni  \scriptfont1=\seveni
\scriptscriptfont1=\fivei
\textfont2=\tensy \scriptfont2=\sevensy
\scriptscriptfont2=\fivesy
\textfont\itfam=\tenit \def\it{\fam\itfam\tenit}
\textfont\mssfam=\tenmss \def\mss{\fam\mssfam\tenmss}
\textfont\bffam=\tenbf \def\bf{\fam\bffam\tenbf} \rm}
\ifx\answ\bigans\def\abstractfont{\tenpoint}\else
\def\abstractfont{\def\rm{\fam0\absrm}
\textfont0=\absrm \scriptfont0=\absrms
\scriptscriptfont0=\absrmss
\textfont1=\absi \scriptfont1=\absis
\scriptscriptfont1=\absiss
\textfont2=\abssy \scriptfont2=\abssys
\scriptscriptfont2=\abssyss
\textfont\itfam=\bigit \def\it{\fam\itfam\bigit}
\textfont\mssfam=\absmss \def\mss{\fam\mssfam\absmss}
\textfont\bffam=\absbf \def\bf{\fam\bffam\absbf}\rm}\fi
%
\def\f@@t{\baselineskip10pt\lineskip0pt\lineskiplimit0pt
\bgroup\aftergroup\@foot\let\next}
\setbox\strutbox=\hbox{\vrule height 8.pt depth 3.5pt width\z@}
\def\vfootnote#1{\insert\footins\bgroup
\baselineskip10pt\footfont
\interlinepenalty=\interfootnotelinepenalty
\floatingpenalty=20000
\splittopskip=\ht\strutbox \boxmaxdepth=\dp\strutbox
\leftskip=24pt \rightskip=\z@skip
\parindent=12pt \parfillskip=0pt plus 1fil
\spaceskip=\z@skip \xspaceskip=\z@skip
\Textindent{$#1$}\footstrut\futurelet\next\fo@t}
\def\Textindent#1{\noindent\llap{#1\enspace}\ignorespaces}
\def\footnote#1{\attach{#1}\vfootnote{#1}}%

\def\foot{\attach\footsymbolgen\vfootnote{\footsymbol}}
\let\footsymbol=\star
\newcount\lastf@@t           \lastf@@t=-1
\newcount\footsymbolcount    \footsymbolcount=0
\def\footsymbolgen{\relax\footsym
\global\lastf@@t=\pageno\footsymbol}
\def\footsym{\ifnum\footsymbolcount<0
\global\footsymbolcount=0\fi
{\iffrontpage \else \advance\lastf@@t by 1 \fi
\ifnum\lastf@@t<\pageno \global\footsymbolcount=0
\else \global\advance\footsymbolcount by 1 \fi }
\ifcase\footsymbolcount \fd@f\star\or
\fd@f\dagger\or \fd@f\ast\or
\fd@f\ddagger\or \fd@f\natural\or
\fd@f\diamond\or \fd@f\bullet\or
\fd@f\nabla\else \fd@f\dagger
\global\footsymbolcount=0 \fi }
\def\fd@f#1{\xdef\footsymbol{#1}}
\def\space@ver#1{\let\@sf=\empty \ifmmode #1\else \ifhmode
\edef\@sf{\spacefactor=\the\spacefactor}
\unskip${}#1$\relax\fi\fi}
\def\attach#1{\space@ver{\strut^{\mkern 2mu #1}}\@sf}
%
\newif\ifnref
\def\rrr#1#2{\relax\ifnref\nref#1{#2}\else\ref#1{#2}\fi}
\def\ldf#1#2{\begingroup\obeylines
\gdef#1{\rrr{#1}{#2}}\endgroup\unskip}
\def\nrf#1{\nreftrue{#1}\nreffalse}
\def\doubref#1#2{\refs{{#1},\ {#2}}}

\nreffalse
\def\refout{\listrefs}
%
\def\eqn#1{\xdef #1{(\secsym\the\meqno)}
\writedef{#1\leftbracket#1}%
\global\advance\meqno by1\eqno#1\eqlabeL#1}
\def\eqnalign#1{\xdef #1{(\secsym\the\meqno)}
\writedef{#1\leftbracket#1}%
\global\advance\meqno by1#1\eqlabeL{#1}}
%
\def\chap#1{\newsec{#1}}
\def\chapter#1{\chap{#1}}
\def\sect#1{\subsec{{ #1}}}
\def\section#1{\sect{#1}}
\def\\{\ifnum\lastpenalty=-10000\relax
\else\hfil\penalty-10000\fi\ignorespaces}
\def\note#1{\leavevmode%
\edef\@@marginsf{\spacefactor=\the\spacefactor\relax}%
\ifdraft\strut\vadjust{%
\hbox to0pt{\hskip\hsize%
\ifx\answ\bigans\hskip.1in\else\hskip-.1in\fi%
\vbox to0pt{\vskip-\dp
\strutbox\sevenbf\baselineskip=8pt plus 1pt minus 1pt%
\ifx\answ\bigans\hsize=.7in\else\hsize=.35in\fi%
\tolerance=5000 \hbadness=5000%
\leftskip=0pt \rightskip=0pt \everypar={}%
\raggedright\parskip=0pt \parindent=0pt%
\vskip-\ht\strutbox\noindent\strut#1\par%
\vss}\hss}}\fi\@@marginsf\kern-.01cm}
\def\titlepage{%
\frontpagetrue\nopagenumbers\abstractfont%
\hsize=\hstitle\rightline{\vbox{\baselineskip=10pt%
{\abstractfont\pubnum}}}\pageno=0}
\frontpagefalse
\def\pubnum{}
\def\pdate{\number\month/\number\yearltd}
\def\makefootline{\iffrontpage\vskip .27truein
\line{\the\footline}
\vskip -.1truein\leftline{\vbox{\baselineskip=10pt%
{\abstractfont\pdate}}}
\else\vskip.5cm\line{\hss \tenrm $-$ \folio\ $-$ \hss}\fi}
\def\title#1{\vskip .7truecm\titlestyle{\titleft #1}}
\def\titlestyle#1{\par\begingroup \interlinepenalty=9999
\leftskip=0.02\hsize plus 0.23\hsize minus 0.02\hsize
\rightskip=\leftskip \parfillskip=0pt
\hyphenpenalty=9000 \exhyphenpenalty=9000
\tolerance=9999 \pretolerance=9000
\spaceskip=0.333em \xspaceskip=0.5em
\noindent #1\par\endgroup }
\def\autskip{\ifx\answ\bigans\vskip.5truecm\else\vskip.1cm\fi}
\def\author#1{\vskip .7in \centerline{#1}}
\def\andauthor#1{\autskip
\centerline{\it and} \autskip\centerline{#1}}
\def\address#1{\ifx\answ\bigans\vskip.2truecm
\else\vskip.1cm\fi{\it \centerline{#1}}}
\def\abstract#1{
\vskip .3in\vfil\centerline
{\bf Abstract}\penalty1000
{{\smallskip\ifx\answ\bigans\leftskip 2pc \rightskip 2pc
\else\leftskip 5pc \rightskip 5pc\fi
\noindent\abstractfont \baselineskip=12pt
{#1} \smallskip}}
\penalty-1000}
\def\endpage{\tenpoint\supereject\global\hsize=\hsbody%
\frontpagefalse\footline={\hss\tenrm\folio\hss}}
\def\ack{\goodbreak\vskip2.cm\noindent {\bf Acknowledgements}}
\def\CERN{\address{CERN, CH--1211 Geneva 23, Switzerland}}
\def\bfone{\relax{\rm 1\kern-.35em 1}}
\def\inbar{\vrule height1.5ex width.4pt depth0pt}
\def\IC{\relax\,\hbox{$\inbar\kern-.3em{\mss C}$}}
\def\ID{\relax{\rm I\kern-.18em D}}
\def\IF{\relax{\rm I\kern-.18em F}}
\def\IH{\relax{\rm I\kern-.18em H}}
\def\II{\relax{\rm I\kern-.17em I}}
\def\IN{\relax{\rm I\kern-.18em N}}
\def\IP{\relax{\rm I\kern-.18em P}}
\def\IQ{\relax\,\hbox{$\inbar\kern-.3em{\rm Q}$}}
\def\IR{\relax{\rm I\kern-.18em R}}
\font\cmss=cmss10 \font\cmsss=cmss10 at 7pt
\def\ZZ{\relax\ifmmode\mathchoice
{\hbox{\cmss Z\kern-.4em Z}}{\hbox{\cmss Z\kern-.4em Z}}
{\lower.9pt\hbox{\cmsss Z\kern-.4em Z}}
{\lower1.2pt\hbox{\cmsss Z\kern-.4em Z}}\else{\cmss Z\kern-.4em
Z}\fi}
\def\nup#1({Nucl.\ Phys.\ $\us {B#1}$\ (}
\def\plt#1({Phys.\ Lett.\ $\us  {#1}$\ (}
\def\cmp#1({Comm.\ Math.\ Phys.\ $\us  {#1}$\ (}
\def\prp#1({Phys.\ Rep.\ $\us  {#1}$\ (}
\def\prl#1({Phys.\ Rev.\ Lett.\ $\us  {#1}$\ (}
\def\prv#1({Phys.\ Rev.\ $\us  {#1}$\ (}
\def\mpl#1({Mod.\ Phys.\ \Let.\ $\us  {#1}$\ (}
\def\tit#1|{{\it #1},\ }
%

%

\def\tilde{\widetilde}
\def\bar{\overline}
\def\us#1{\underline{#1}}

\def\Coe#1.#2.{{#1\over #2}}
\def\coeff#1#2{\relax{\textstyle {#1 \over #2}}\displaystyle}
\def\coe#1.#2.{\relax{\textstyle {#1 \over #2}}\displaystyle}

\def\notin{\hbox{{$\in$}\kern-.51em\hbox{/}}}

\def\del{\partial}

 \def\ie{{i.e.}}
\catcode`\@=12
%

\def\chargino{\chi^{\pm}}

\def\mhiggs{m_h}
\def\mqsi{m_{q_i}}
\def\coi{l_i}
\def\cti{n_i}
\def\QL{Q_L}

\def\UR{U_R}

\def\DR{D_R}

\def\LL{L_L}

\def\ER{E_R}

\def\hone{H_1}
\def\htwo{H_2}
\def\hn{h^0}

\def\ytop{\yuk_{t}}
\def\yuk{y}
\def\mz{M_{Z}}

\def\mtop{m_t}
\def\mg{\tilde m}
\def\ind{a}
\def\hc{\rm h.c.}
\def\del{\partial}

\def\Weff{W^{\rm (eff)}}
\def\Veff{V^{\rm (eff)}}

\def\mpl{M_{\rm Pl}}
\def\mgut{M_{\rm X}}

\def\alphagut{\alpha_{\rm X}}
\def\mgrav{m_{3/2}}

\def\mb{B}
\def\mur{\mu_R}

\def\matter{Q}

\def\smatter{q}
\def\smatterb{
\smash{\overline\smatter}\vphantom{\smatter}}

\def\Jb{{\bar J}}

\interdisplaylinepenalty=10000
\def\K{K\"ahler}
\def\brk{\hfill\break}
%
%
\ldf\MSSM{For a review see for example,\brk
H.-P.~Nilles, \prp C110 (1984) 1;\brk
H.E.~Haber and G.~Kane, \prp C117 (1985) 75;\brk
R.~Barbieri, Riv. Nuovo Cimento $\us{11}$ (1988) 1;\brk
L.E.~Ib\'a\~nez and G.G.~Ross, CERN preprint CERN-TH.6412/92,
 to appear in {\it Perspectives in Higgs Physics}, ed.~ G.~Kane;\brk
F.~Zwirner,  preprint
CERN-TH.6357/91, Talk at the Workshop on Physics and Experiments with
Linear
Colliders, Saariselka, Finland, Sep.~1991;
and references therein.
}
\ldf\gauginoc{For a review see for example,\brk
 D. Amati, K. Konishi, Y. Meurice, G. Rossi and
G. Veneziano, \prp  162 (1988) 169;\brk
 H.-P.~Nilles,
Int. J. Mod. Phys. $\us{A5}$ (1990) 4199;\brk
J. Louis, in  Proceedings of the 1991 DPF meeting, World
Scientific, 1992;\brk
 and references therein.}
\ldf\GM{G.~F.~Giudice and A.~Masiero, \plt 206B (1988) 480.}
\ldf\KN{J.E.~Kim and H.-P.~Nilles, \plt 138B (1984) 150;\brk
J.L.~Lopez and D.V.~Nanopoulos, \plt B251 (1990) 73;\brk
E.J.~Chun,  J.E.~Kim and H.-P.~Nilles, \nup370 (1992) 105. }
\ldf\FILQ{
A.~Font, L.E.~Ib\'a\~nez, D.~L\"ust and F.~Quevedo,
\plt  B245 (1990) 401.}
\ldf\CFILQ{M.~Cveti\v c, A.~Font, L.E.~Ib\'a\~nez, D.~L\"ust and
F.~Quevedo, \nup361 (1991) 194.}
\ldf\IL{L.~Ib\'a\~nez and D.~L\"ust, \nup382 (1992) 305.}
\ldf\krasnikov{N.V.~Krasnikov, \plt  193B (1987) 37;\brk
L.~Dixon, in {\it Proceedings of  the A.P.S. D.P.F. Meeting},
Houston, 1990;\brk
J.~A.~Casas, Z.~Lalak, C.~Mu\~noz and G.G.~Ross, \nup347 (1990)
243;\brk
T.~Taylor, \plt B252 (1990) 59.}
\ldf\KLb{V.~Kaplunovsky and J.~Louis, \plt B306 (1993) 269.}
\ldf\ILM{
 L.E.~Ib\'a\~nez and  C.~L\'opez, \nup233 (1984) 511;\brk
 L.E.~Ib\'a\~nez, C.~L\'opez and C.~Mu\~noz, \nup256 (1985) 218.}
\ldf\BG{R.~Barbieri and G.F.~Giudice, \nup306  (1988) 63.}
\ldf\cvetictalk{M.~Cveti\v c, University of Pennsylvania preprint
UPR-528-T.}
\ldf\gauginoall{J.P.~Derendinger, L.E.~Ib\'a\~nez and
  H.P.~Nilles, \plt  155B (1985) 65;\brk
M.~Dine, R.~Rohm, N.~Seiberg and
  E.~Witten, \plt  156B (1985) 55;\brk
A.~Font, L.E.~Ib\'a\~nez, D.~L\"ust and F.~Quevedo,
\plt  B245 (1990) 401;\brk
S.~Ferrara, N.~Magnoli, T.~Taylor and G.~Veneziano, \plt
 B245 (1990) 409;\brk
H.P.~Nilles and M.~Olechowski, \plt B248 (1990)
268;\brk
P.~Bin\'etruy and M.K. Gaillard, \nup358 (1991) 121;\brk
B.~de Carlos, J.A.~Casas and C.~Mu\~noz, preprint
CERN-TH.6436/92, \plt B299 (1993) 234;\brk
V.~Kaplunovsky and J.~Louis,  to appear.}
\ldf\higgsref{Y.~Okada, M.~Yamaguchi and T.~Yanagida,
Prog.~Theor. \plt 85 (1991) 1;\brk
J.~Ellis, G.~Ridolfi and F.~Zwirner, \plt B257 (1991) 83;\brk
H.E.~Haber and R.~Hempfling, \prl 66 (1991) 1815;\brk
R.~Barbieri, M.~Frigeni and M.~Caravaglios, \plt B258 (1991) 167.}
\ldf\MR{A.~de la Macorra and G.G.~Ross,
Oxford preprint PRINT-92-0463.}
\ldf\neutralinolimit{J.-F.~Grivaz, Proceedings of the Rencontres de
Moriond, Les Arcs, France, March 1993.}
\ldf\toplimit{B.~Harral, Proceedings of the Rencontres de Moriond,
Les Arcs, France, March 1993.}
\ldf\higgslimit{T.~Mori, Proceedings of the XXVI International
Conference on High Energy Physics, Dallas, 1992.}
\ldf\FLST{S.~Ferrara, D.~L\"ust, A.~Shapere and
  S.~Theisen, \plt  B225 (1989) 363.}
\ldf\ILR{J.~Ellis, S.~Kelley and D.V.~Nanopoulos, \plt249 (1990) 441;\brk
L.~Ib\'a\~nez, D.~L\"ust and G.~Ross, \plt272 (1991) 251.}
\ldf\AEKN{I.~Antoniadis, J.~Ellis, S.~Kelley and D.V.~Nanopoulos,
\plt B272 (1991) 31.}
%
%

\def\pubnum{
\hbox{CERN-TH.6856/93}
}
\def\pdate{
\hbox{CERN-TH.6856/93}
\hbox{April 1993}
}
\titlepage
\title
{Phenomenological Implications of
Supersymmetry Breaking by the Dilaton}
\vskip -1.0cm
\author{
Riccardo Barbieri}
\address{Department of Physics, University of Pisa}
\address{INFN sez. di Pisa, Italy}
\centerline{{\it and}}
\vskip -0.2cm
\CERN
\vskip -1.0cm
\author{Jan Louis}
\CERN
\andauthor{
Mauro Moretti}
\address{SISSA, Trieste; INFN sez. di Trieste }
\address{
 and INFN sez. di Ferrara, Italy}
\vskip 1.0truecm
{\centerline {\bf Abstract}}

\noindent
We investigate the low energy properties of  string vacua
with spontaneously broken $N=1$ supersymmetry by a
dilaton $F$-term. As a consequence of the universal couplings of the
dilaton,
  the supersymmetric mass spectrum is  determined
in terms of only three independent parameters
and    more constrained than in the minimal supersymmetric Standard
Model.
For a $\mu$-term induced by  the
\K\ potential the parameter space
becomes two-dimensional; in the allowed regions of this parameter
space we find that most supersymmetric particles are
 determined solely by the
gluino mass. The      Higgs is rather light
and the top-quark mass always lower than 180 GeV.

\endpage

The leading candidate for consistently incorporating quantum gravity
into the standard  interactions of particle physics
is a heterotic superstring theory.
However,
despite its numerous attractions
it has not been possible
 to
identify  quantitatively
the Standard Model (SM) as the low energy limit
of string theory.
In part this is due to our
lack of
conceptual
understanding of the theory;  so far we only enjoy control over (some
of)
its  perturbative regime. Unfortunately,
low energy
string phenomenology does seem to depend crucially on
non-perturbative
properties of string theory.
The mechanism for supersymmetry
breaking, the choice of the string vacuum, or the determination of
the gauge couplings
are believed to be governed by (possibly `stringy') non-perturbative
effects; our current techniques are inappropriate to incorporate such
effects into the low
 energy effective Lagrangian.

Ultimately, we have to come to terms with this deficiency;
in the mean time
various strategies have been employed in order to investigate
and/or constrain
the low energy limit of the string.
We refrain here from systematically reviewing the subject,
instead we briefly outline the method that we are going to follow
in this letter.
Based on  work in the context of gaugino condensation
\doubref\gauginoall\IL\ and  duality-invariant
effective Lagrangians
\nrf{\FLST\CFILQ}
\refs{\IL {--} \CFILQ}
 it was recently suggested \KLb\ to
simply parametrize the unknown
non-perturbative physics.
All relevant
low energy interactions are  expressed in terms of couplings
calculable in string perturbation theory
and couplings
encoding the non-perturbative dynamics;
the latter
 then appear as arbitrary parameters in the
low energy effective theory.
Surprisingly, even in such a general framework
this effective theory can display rather
distinct properties.
In ref.~\KLb\  the non-perturbative couplings
are constrained
by some
assumptions about the nature of the non-perturbative dynamics
and the nature of supersymmetry
breaking. In particular,
it was assumed that supersymmetry is spontaneously
broken in the moduli/dilaton sector of
string theory.
Such scalar multiplets are always present in
the massless string spectrum
and
the couplings of the low energy effective Lagrangian are determined
by their vacuum expectation values
(VEVs).
In string perturbation theory, both the moduli and the dilaton
 are exact
flat directions of the effective potential, leaving their VEVs
undetermined.
In ref.~\KLb\  this
perturbative degeneracy is assumed to be completely lifted
by  the non-perturbative dynamics
and    VEVs for moduli and dilaton
to be induced. In addition,
 supersymmetry is assumed to be
spontaneously broken by the auxiliary $F$-terms of the
moduli/dilaton  supermultiplets.
Indeed, in the context of gaugino condensation, such
a scenario can  occur \doubref\gauginoall\krasnikov.

The dilaton plays a distinct
role in the low energy theory;
all  its  couplings  are
universal (at the tree level), that is,
  they are identical for all $N=1$ heterotic string
vacua or equivalently they do not depend
on the details of the internal conformal
field theory.\foot{The dilaton VEV also determines the tree-level
gauge couplings.}
As a consequence supersymmetry breaking dominated by  the dilaton
$F$-term leads
to very specific and model-independent
low energy properties.
It is the purpose of this letter to study the phenomenological
implications of  supersymmetry breaking in the dilaton sector.
Such an analysis
has not been done previously since supersymmetry, in the context of
gaugino
condensation,  usually
 breaks  in the moduli direction.\foot{
See however ref.~\MR.}
  However, if one does not specify the
non-perturbative physics  (in the spirit of ref.~\KLb),
supersymmetry breaking by a dilaton $F$-term is a conceivable
scenario. As we will see from a phenomenological point of view it
automatically leads to some desired features.

 In string perturbation theory
there is an enormous vacuum degeneracy and out of this plethora
of possibilities  one chooses (by hand)
phenomenologically promising  candidate vacua. For the purpose of
this article we
require
the class of string vacua under consideration to satisfy
a few standard  properties.
In addition to the moduli and dilaton,
the string spectrum contains families of matter multiplets that are
charged under the gauge group $G$.
Part of this gauge group
has to contain  the standard $SU(3)\times SU(2) \times U(1)$,
and we
denote all light $N=1$
chiral multiplets in this `observable sector'
by $\matter^I$.
For simplicity we assume that the $\matter^I$  coincide with
the
multiplets present in the minimal supersymmetric
Standard Model
(MSSM), that is, all particles
of the SM occur in  chiral superfields with one additional Higgs
doublet  \MSSM. (This assumption is not crucial for the  structure
of the soft supersymmetry-breaking terms themselves;
 however, some of the low energy
properties
do depend for example on the Higgs sector.)
Thus, the index $I$ labels collectively
the
quark multiplets ($\QL, \UR, \DR$), the leptons ($\LL, \ER$)
 and  the two Higgs doublets ($\hone, \htwo$),
and we suppress their gauge quantum numbers.\foot{Note that
$\QL$ denotes the left-handed quark supermultiplet and should not be
confused with $\matter^I$, which stands for all matter multiplets.}

The low energy
interactions of the observable fields
consist of  supersymmetric couplings
(encoded in a superpotential $W$) and a set of soft
supersymmetry-breaking
parameters.
The masses and Yukawa couplings of the chiral (matter)
 fermions are summarized by an effective superpotential of the  form
$$
\Weff =  \coeff12 \mu_{IJ}\
\matter^I \matter^J\, + \coeff13 \ \yuk_{IJK}\ \matter^I
\matter^J \matter^K\, ,
\eqn\weffdef
$$
where we have chosen  a basis for $\matter^I$ with
canonically normalized
 kinetic energy terms.
Thus, $\mu_{IJ}$ are the physical (supersymmetric) mass terms and
$\yuk_{IJK}$
 denote the physical Yukawa couplings.\foot{
Our definition of $\matter^I, \mu_{IJ}$ and $\yuk_{IJK}$ differs
from the definition in eqs.~(9) and (10) of  \KLb\
in that here we use canonically normalized fields throughout.}
 These
can be computed as
 functions of the moduli in
string perturbation theory, and we assume here that they reproduce
the known Yukawa couplings of the MSSM. On the other hand $\mu_{IJ}$
is generated
either in string perturbation theory or by non-perturbative
effects \doubref\GM\KN;
because of  gauge invariance it has only one non-vanishing entry in
the
direction
of the two Higgs doublets ($\mu_{12} \equiv \mu$).
Thus, in the standard  notation of the MSSM, eq.~\weffdef\ reads
$$
\Weff=  \mu \hone \htwo
 + \sum_{\rm generations} \left( \yuk_{U} \QL \UR \htwo
+ \yuk_{D} \QL \DR \hone + \yuk_{L} \LL \ER \hone \right)  \, .
\eqn\weffmssm
$$

In addition to the supersymmetric interactions (eqs.~\weffdef,
\weffmssm)
supersymmetry breaking induces  soft
breaking parameters in the observable sector.
The general structure of these soft terms in string theory
was analysed in
refs.~\refs{\CFILQ{, }\IL{, }\KLb}\ and will not be repeated here.
Instead, we just recall their form  for the particular case of
supersymmetry
breaking induced by
a dilaton  $F$-term.
Because of the universal couplings of the dilaton,
the entire effect of the  breaking
can be
parametrized by the gravitino mass $\mgrav$.
Neglecting string loop corrections
one finds  a universal (gauge-group-independent)
gaugino mass, which is
determined by $\mgrav$\foot{
The gaugino mass  given in ref.~\KLb\ (eqs.~(8) and (15))
incorrectly  includes a factor of $\coeff12$.  We thank L.~Ib\'a\~nez
for pointing  this out.}
$$
\mg_a\ =\ \sqrt 3\ \mgrav \, , \quad \forall  a \ ,
\eqn\gauginomass
$$
where we label the different factors in the
observable gauge group $G$  according to
$G = \prod_\ind G_\ind$.
(The  mass term of eq.~\gauginomass\ is given in a basis where the
gauginos are  canonically normalized.)
The potential for the scalar fields $\smatter^I$ in the
supersymmetric multiplets  takes the form
$$
\eqalign{
\Veff(\smatter,\smatterb) \ ={}&\
\sum_{\ind}{g_\ind^2\over 4}
    \left( \smatterb  T_\ind \smatter\right)^2\
+\ |\del_I\Weff|^2  \cr
+&\ m^2_{I\Jb}\smatter^I \smatterb^\Jb\
+\ \left(\coeff13 A_{IJL}\smatter^I\smatter^J\smatter^L\,
    +\,\coeff12\mb_{IJ}\smatter^I\smatter^J\ +\ \hbox{h.c.}\right)\,
{}.
}
\eqn\Vsoft
$$
The first two terms are the standard supersymmetric
potential,  whereas the last three are soft
supersymmetry-breaking interactions.
Their structure for dilaton-induced supersymmetry breaking is highly
constrained and given by \KLb (again neglecting string loops)
$$
m^2_{I\Jb}\ =\ \mgrav^2\ \delta_{I\Jb}\,,\qquad
A_{IJL}\ =\ -\sqrt{3}\ \mgrav\  \yuk_{IJL}\,;
\eqn\sdom
$$
 $\mb_{IJ}$ also has only one non-vanishing entry --
 the coefficient of the Higgs doublets
 ($\mb_{12} \equiv \mb$) -- but in general is not restricted further.
{}From eq.~\sdom\ we learn  that all scalar masses
$m^2_{I\Jb}$
are
flavour-independent (universal) and furthermore the
$A$-terms are strictly proportional to the Yukawa couplings with a
universal
 constant of proportionality.
Both features  are commonly
assumed in phenomenological investigations of the MSSM, but
generically do not
 hold
 in string theory.
As a consequence, `dilaton breaking'
automatically
ensures the smallness of flavour-changing neutral
currents (FCNC).
This is not guaranteed
in other scenarios of supersymmetry breaking and
in general
 imposes  strong constraints on
the perturbative couplings of the string vacuum \IL.
The other distinct feature displayed by
eqs.~\gauginomass\ and \sdom\ is the fact
that gaugino and scalar  masses, as well as $A$-terms, are
locked
 in terms of
$\mgrav$ with no free parameter to vary.
This leads to significant constrains on the low-energy mass spectrum
and  we find part of this spectrum directly determined by $\mgrav$.
To summarize, the entire supersymmetric mass spectrum is  expressed
in terms  of only
three independent parameters
$\mgrav$, $\mu$ and $\mb$.\foot{
Note that
$\mb$ is  not necessarily proportional
to $\mu$.}

This three-dimensional parameter space can be further reduced
 if one specifies the mechanism responsible for
generating the
$\mu$-term. Generically, there is a danger in string theory
of inducing
a large $\mu$, which
prohibits a light Higgs. However, if
$\mu$ arises from couplings in the \K\ potential  (which do occur
in string theory)
its size is  automatically
$O(\mgrav)$ \GM.
For a $\mu$-term solely generated by this mechanism,\foot{
This corresponds to $\tilde\mu = 0$ in eqs.~(2) and (9) of
ref.~\KLb.}
 $\mb$ is no longer an independent parameter but instead obeys
$\mb=2\, \mu\, \mgrav$ \KLb.
In this case the mass spectrum is determined by two independent
parameters, $\mgrav$ and $\mu$.

The mass relations of eqs.~\gauginomass\ and \sdom\ should be viewed
as a boundary
condition
at the unification scale $\mgut$ before (low-energy)
renormalization effects are taken into
account.
The mass spectrum of the supersymmetric
particles at the weak scale is determined by the
 evolution    of
the  couplings according to their
renormalization group (RG) equations.
 Here, we use the standard RG analysis where only the top-quark
Yukawa coupling $\ytop$
is kept  \ILM.\foot{
Our numerical calculation also takes into account the effects of the
supersymmetry threshold.}
As the unification scale we choose
$\mgut = 3 \times 10^{16}$ GeV in order to be consistent with the
unification of the gauge couplings.
String theory indeed implies a unification of  gauge couplings;
 however, it occurs at  the characteristic string
scale, which is  approximately $5\times 10^{17}$
 GeV and  does not
coincide with $\mgut$.
 There exist various suggestions of how to remedy this fact and
 we assume here that
 the
string scale is effectively lowered by large threshold corrections
\doubref\ILR\cvetictalk.\foot{
Another possibility would be to assume  extra light states
in the spectrum, which decouple at some intermediate scale \AEKN, but
we do not entertain this option here.}

Let us turn  to the Higgs sector, which is
responsible for the electroweak symmetry breaking \MSSM.
{}From eq.~\Vsoft\ we learn that  the  potential for
the two neutral components $\hn_1, \hn_2$ of the
 Higgs doublets is given by
$$
V = \coeff18 (g^2_1 + g^2_2) (|\hn_1|^2 - |\hn_2|^2)^2 + m_1^2
|\hn_1|^2 + m_2^2 |\hn_2|^2 - m_3^2 (\hn_1 \hn_2 +\hc) \, ,
\eqn\higgspot
$$
with the boundary conditions at $\mgut$
$$
m_1^2 = m_2^2 = \mgrav^2 + \mu^2 \, ,
\qquad m_3^2 = - B\, .
\eqn\rel
$$
These mass parameters
evolve according to their RG equation,
which can be solved analytically, as a function of  the top Yukawa
coupling
$\ytop$ only. At  low energies one finds
(assuming \gauginomass\ and \sdom\ to hold) \ILM
$$
\eqalign{
m_{1}^2 &=
  c_{1}\, \mgrav^2 +\  \mur^2\, ,\qquad
m_{2}^2 =
  c_{2}(\ytop)\, \mgrav^2 +\ \mur^2 \, ,\cr
m_3^2 &=  c_4(\ytop)\, B +\ c_5(\ytop)\, \mur\, \mgrav\, ,\qquad
\mur^2 = c_3(\ytop)\, \mu^2\, ,
}
\eqn\mrenormalized
$$
where $c_{2-5}$ develop a (complicated)  dependence on the unknown
$\ytop$;
 their precise functional form can be found in ref.~\ILM.
For the following analysis we only need to record that
 $c_1$ is independent of $\ytop$ whereas $c_2$ obeys  $c_2(\ytop) \le
c_1$ and reaches its maximum at $\ytop =0$, \ie\ $c_2(0)= c_1$.
Furthermore, at low energies the renormalized $\ytop$ cannot grow
arbitrarily, but is instead  `attracted' by an infrared fixed point
of
its RG equation $\ytop \rightarrow \ytop^{\rm crit}$. At that fixed
point the coefficient $c_3$  vanishes: $c_3(\ytop = \ytop^{\rm crit})
= 0$.

In order to induce  electroweak symmetry breaking
 the renormalized masses have to satisfy
$$
2 m_3^2 < m_1^2 + m_2^2\, ,  \qquad  m_1^2 m_2^2 < m_3^4\, ,
\eqn\stability
$$
and
$$
\mz^2 = 2\ {m_1^2 - m_2^2\ \tan^2\beta \over \tan^2\beta - 1}\, ,
\eqn\mzconstrain
$$
where
$$
\tan\beta ={ \vev{\hn_2} \over \vev{\hn_1}}\, , \qquad
\sin 2\beta = {2 m_3^2\over m_1^2 + m_2^2}\, ,
\qquad {\pi \over 4} \le \beta \le {\pi\over 2}\, .
\eqn\sintan
$$
The relations \stability\
do not
 hold automatically but constrain  the initial soft parameter space
as we will see shortly.
Even though $\ytop$ is unknown, it cannot be viewed as an independent
parameter because of  the constraint equation \mzconstrain.
Which parameter one chooses to eliminate via \mzconstrain\  is a
matter of convenience
and taste.
In our analysis,  we  eliminate $\ytop$ and determine the top-quark
mass $\mtop\ (= \ytop\, \vev{\hn_2})$ as a function of the soft
parameters.
Equivalently, one could trade $\ytop$ for one of the soft parameters
and use instead $\mtop$ as an input parameter.

One of the distinct features of the  supersymmetric Higgs potential
\higgspot\ is
the occurrence of a light  Higgs boson. At the tree level its mass is
given by
$$
\eqalign{
\mhiggs^2\ =\ \coeff12 \left[ m_A^2 + \mz^2  -
\left((m_A^2 + \mz^2)^2 - 4 m_A^2 \mz^2 \cos^2
2\beta\right)^{\coeff12}\right]\ ,\cr
{\rm where} \quad m_A^2= m_1^2 + m_2^2\ ,}
\eqn\higgsmass
$$
and  $m_1, m_2$ are as defined in
eq.~\mrenormalized. In the limit $\tan\beta \rightarrow 1$, $\mhiggs$
approaches zero.\foot{It has recently been realized that
  one-loop corrections can significantly
raise $\mhiggs$, because of the heavy top quark \higgsref. In our
numerical evaluation, we take this
into account.}

After these preliminaries we are in a position to discuss the
mass spectrum of the supersymmetric particles.
We start our investigation with the three-parameter case and
afterwards
consider  the two-parameter scenario.
We confine our attention to those features of the mass spectrum that
differ from the
 typical MSSM results.
At an arbitrary RG scale $p$,
the gaugino masses obey
$$
\mg_a(p) = k_a\ {\alpha_a (p)\over \alphagut}\ \mg(\mgut)\, ,
\qquad
k_2 =k_3=1\, , \quad k_1 = 5/3\, .
\eqn\gmassevol
$$
Equation \gauginomass, together with
$\alpha_3(\mz)\approx 0.118$ and
$\alphagut\approx 1/24$, implies  at low energies
$$
\mg_a = d_a\,  \mgrav \, ; \qquad
d_3 \approx 5\, , \quad d_2 \approx 1.5\, ,\quad d_1 \approx 0.75\ .
\eqn\gluino
$$
($\mg_3$ is the gluino and should not be confused with $m_3$ of
eq.~\mrenormalized.)
Again because of eq.~\gauginomass\
all squark and slepton masses $\mqsi$ (except the stop mass)  are
 determined by $\mgrav$
(or equivalently $\mg_3$). One finds \ILM
$$
\mqsi^2 = \coi\,  \mg_3^2\ +\ \cti\, \mz^2 \, ,
\quad {\rm where}\quad
\coi \ge 0.3\, , \quad
-\coeff12<\cti<\coeff12\, .
\eqn\ssmass
$$
The $\coi$ are  fixed numerical coefficients
with no dependence on the soft parameters, whereas
$\cti$ depend on $\tan\beta$.
Instead of listing $\coi$ and $\cti$,
 we  display the squark and lepton masses as a function of the gluino
mass $\mg_3$ in  fig.~1.
Owing to the small  second term in
eq.~\ssmass,
 they lie within a tiny band, which is invisible in fig.~1. The
slepton masses coincide to a very good approximation
with
$0.3\,\mg_3$, whereas the squark masses are essentially degenerate with
$\mg_3$.
Because of the (large) top Yukawa coupling, the left- and
right-handed stop
can have a large mixing term; as a consequence the stop mass is
not accurately described by eq.~\ssmass. For large $\mu$ it can be
significantly lower than the other squark masses \MSSM.

The masses of the four neutralinos $\chi^0$
(linear combinations of Higgsinos, photino and zino)
are determined by the eigenvalues of a $4\times 4$ mass matrix with
input parameters  $\mg_3,\ \mur$ and
$\tan\beta$  \MSSM.
 The scale of the lowest eigenvalue is set either  by
$\mg_1\ (\approx 0.16\, \mg_3)$ or by $\mur$, whichever is lower.
Since the lightest slepton mass is approximately
$0.3\,\mg_3$
we immediately conclude
that the lightest supersymmetric particle (LSP) is always a
neutralino.
Its  mass range is similar to the mass range found in the standard
MSSM analysis, which can be understood from the fact that in both
cases the neutralino masses are determined by three independent
parameters.
The masses of the charginos $\chargino$
(linear combination of the charged Higgsino and the charged wino) are
determined by the exact
same three input parameters
and, as a consequence, we find no significant deviation from the MSSM
in the
chargino sector.
A similar conclusion holds in the Higgs sector.
In our numerical analysis  we find no
restriction of $\tan\beta$ and consequently $\mhiggs$ also
varies  within the standard ranges over the allowed parameter space.
This will
change in the two-parameter case to which  we
now
turn our attention.

The supersymmetric mass spectrum only depends on two soft parameters
if  $\mu$ is
generated by terms in the \K\ potential as was first suggested
 in ref.~\GM. In this case
$\mb$ is no longer  independent but  obeys
$\mb=2\, \mu\, \mgrav$.
With respect to the three-dimensional parameter space we just
discussed, the  main difference arises from  the fact
that $\mu$ is now constrained to lie well above $\mgrav$.
Numerically we find that \foot{
The mass spectrum is symmetric under the exchange of $\mu \rightarrow
- \mu$ and independently under $\mgrav \rightarrow - \mgrav$ and
therefore we only consider $\mu \ge 0$ and $\mgrav \ge 0$.}
$$
\mu >
0.4\, \mg_3
\qquad {\rm and}
\qquad
\mg_3 > 225\, {\rm GeV}
\eqn\mulimit
$$
has to be satisfied  in order to evade the experimental bounds on the
top-quark mass ($\mtop \ge 108$ GeV \toplimit)
and the Higgs mass \higgslimit.
In deriving \mulimit\
 we  first observe
 that the top mass limit alone pushes
 the mass of  the pseudoscalar Higgs $m_A$  well above
70~GeV
(this
can be seen from \mrenormalized\ and the second equation in
\higgsmass) and
as a consequence the scalar Higgs becomes effectively
the SM Higgs with a lower bound of $\approx 60$ GeV \higgslimit.
The combined limits of $m_t\ge 108$ GeV and $m_h\ge 60$~GeV
then result in the constraint \mulimit.
 (For gluino masses close to their lower bound $\mu$ has to be  bigger
than $\mg_3$.)
In Fig.~2 we display the top and Higgs mass
(along with the lightest neutralino and chargino)
as a function of $\mu$ for a fixed gluino mass $\mg_3 = 400$~GeV.
We clearly see that the experimental bounds imply  a large  $\mu$.
Analytically,
the constraint \mulimit\ can be  understood  from the fact
that for
$\mu=\mgrav$ at $\mgut$ the Higgs mass matrix
(which we can read off from eq.~\higgspot) has a zero eigenvalue
and (almost) causes an instability after RG effects are included.
In addition, for small $\mu$ eqs.~\mzconstrain\ and \mrenormalized\
force $\ytop$ to very small values, which results in a low
top-quark mass.

For large $\mu$ we find a rather different behaviour. From
eqs.~\mzconstrain\ and \sintan\ we immediately infer that  a large $\mu$
is
only accessible if at the same time $c_3 \rightarrow 0$, such that
$\mur$ stays fixed. As we already indicated, this behaviour of $c_3$
precisely occurs for $\ytop$ approaching its infrared fixed point.
Since  the physical masses  depend on $\mur$,
they display an asymptotic behaviour as a function of
$\mu$ which can be observed in Fig.~2.
As a consequence,
 the lightest neutralino, chargino and stop as well as
 the pseudoscalar Higgs effectively depend only
$\mg_3$ in the allowed region of large $\mu$.
Fig.~3 displays their masses as a function of $\mg_3$;
the stop and the chargino retain a weak dependence on $\mu$
which results in the `spread' seen in the plot.
The
squark
and slepton masses (except the stop) again obey eq.~\ssmass,
where $\cti$ is now  a
pure
number. Since this second term is small,  Fig.~1 adequately
summarizes
the squark masses also in this case;
 the stop is the lightest squark with a  mass   well above $\mtop$.
Thus, for  most of the supersymmetric particles  the original
two-dimensional parameter space effectively reduces to a one-dimensional
space with the masses determined solely by $\mg_3$.

For the Higgs and the top quark the parameter space remains two-dimensional.
 However, from eqs.~\mzconstrain\ and \mrenormalized\ one infers that
 for large $\mu$, $\tan\beta$ is almost independent of $\mg_3$
which results in an upper bound  $\tan\beta \le 2$.
 This in turn, implies an upper  bound on the top mass
$\mtop \le 180$ GeV and leads to a relatively light Higgs boson
all over the allowed parameter space
(as can be  seen from eq.~\higgsmass).
In Fig.~4 we show the (one-loop corrected) Higgs mass as a
 function of $m_t$  for different values of $\mg_3$.
 We see that both top and Higgs are constrained and
only for a  large gluino  and a large top mass the
Higgs can be heavier than the $Z$-boson.
In addition, for  fixed gluino mass  there
is a linear correlation between Higgs and top mass.


Let us conclude.
We investigated the low-energy supersymmetric mass spectrum, which
arises
under the assumption that supersymmetry is spontaneously
broken by the dilaton
$F$-term. Owing to the universal couplings of the dilaton, the
structure
of the soft parameters  as given in eqs.~\gauginomass\ and \sdom\
holds for all $N=1$ vacua of the heterotic string.
Compared with other scenarios of supersymmetry breaking in string
theory, they display some simplicity and have a phenomenological
appeal. The scalar and gaugino masses as well as the $A$-terms
are automatically universal (a feature that generically  does not
hold in string theory)
and determined in terms of $\mgrav$.
Without specifying the
 mechanism for generating the $\mu$-term,
  the supersymmetric mass spectrum is  determined
in terms of only three independent parameters $\mgrav,\, \mu$, and
$B$,
and  as a consequence
the masses are slightly more constrained than in the MSSM.
For a $\mu$-term induced by the
\K\ potential \GM, $B$ is related to $\mu$ and the parameter space
becomes two-dimensional. Current
experimental limits of the Higgs and the top-quark mass
 further constrain the range of $\mu$
and  lead to a reduction of the parameter space for
 most of the supersymmetric particles. $\tan\beta$
 is always small and  as a consequence the Higgs is
rather light whereas the top  mass is bounded by  180~GeV.

\ack

\noindent
We would like to thank G.~Giudice, V.~Kaplunovsky and F.~Zwirner
for helpful conversations and M.~Dine and D.~L\"ust for useful
comments on the manuscript.
\vskip 1cm
\noindent
{\bf Note added}

In a previous version of this article the gaugino masses
(eq.~\gauginomass) were given incorrectly. We thank L.~Ib\'a\~nez
for correcting this error.

After completion of this paper we received a
preprint by J.~Lopez, D.~Nanopoulos and A.~Zichichi,
(CERN-TH.6903/93),
 which analyses supersymmetry breaking by the dilaton
 in the context of a flipped $SU(5)$ model.
\refout
\endpage
%
%
{\bf Figure captions}
\vskip 1.0cm
\item{Fig.~1}  Masses of squarks (except stop)  $Q_L$ (---), $U_R,
D_R$ (- - -), and sleptons $L_L$ $(- \cdot - \cdot -)$, $E_R$
($\cdots$)
as a function of the gluino mass
$\mg_3$.
\vskip 0.5cm
\item{Fig.~2} Masses of top quark (---), lightest scalar Higgs boson (- - -),
 lightest
neutralino $(- \cdot - \cdot -)$
and lightest chargino ($\cdots$)
as a function of $\mu$ for
a gluino mass $\mg_3 = 400\, GeV$.
\vskip 0.5cm
\item{Fig.~3} Masses of pseudoscalar Higgs ($---$),
lightest stop (---),  lightest chargino ($---$) and
 lightest
neutralino $(- \cdot - \cdot -)$
for large $\mu$ as a function of the gluino mass $\mg_3$.
\vskip 0.5cm
\item{Fig.~4}  Higgs mass as a function of top-quark mass
 for gluino masses \hfill\break
$\mg_3 = 250, 500, 1000, 1500$~GeV.
\vfill
\end